\documentclass[twocolumn,showpacs,preprintnumbers,amsmath,amssymb,superscriptaddress,aps,prb]{revtex4-1}

\usepackage{graphicx}
\usepackage{dcolumn}
\usepackage{bm}
\usepackage{color}

\begin{document}


\title{The interface between heavy fermions and normal electrons investigated by spatially-resolved nuclear magnetic resonance}

\author{Takayoshi Yamanaka}
 \affiliation{Department of Physics, Kyoto University, Kyoto 606-8502, Japan}
 \email{t-yamanaka@scphys.kyoto-u.ac.jp}
\author{Masaaki Shimozawa}
 \affiliation{Institute for Solid State Physics, University of Tokyo, Kashiwa, Chiba 277-8581, Japan}
\author{Ryota Endo}
 \affiliation{Department of Physics, Kyoto University, Kyoto 606-8502, Japan}
\author{Yuta Mizukami}
 \affiliation{Department of Advanced Materials Science, University of Tokyo, Kashiwa, Chiba 277-8561, Japan}
\author{Hiroaki Shishido}
 \affiliation{Department of Physics and Electronics, Osaka Prefecture University, Sakai, Osaka 599-8531, Japan}
\author{Takahito Terashima}
 \affiliation{Research Center for Low Temperature and Materials Sciences, Kyoto University, Kyoto 606-8501, Japan}
\author{Takasada Shibauchi}
 \affiliation{Department of Physics, Kyoto University, Kyoto 606-8502, Japan}
 \affiliation{Department of Advanced Materials Science, University of Tokyo, Kashiwa, Chiba 277-8561, Japan}
\author{Yuji Matsuda}
 \affiliation{Department of Physics, Kyoto University, Kyoto 606-8502, Japan}
\author{Kenji Ishida}
 \affiliation{Department of Physics, Kyoto University, Kyoto 606-8502, Japan}
 \email{kishida@scphys.kyoto-u.ac.jp}
\date{\today}

\begin{abstract}
We have studied the superlattices with alternating block layers (BLs) of heavy-fermion superconductor CeCoIn$_5$ and conventional-metal YbCoIn$_5$ by site-selective nuclear magnetic resonance(NMR) spectroscopy, which uniquely offers spatially-resolved dynamical magnetic information. 
We find that the presence of antiferromagnetic fluctuations is confined to the Ce-BLs, indicating that magnetic degrees of freedom of $f$-electrons are quenched inside the Yb-BLs. 
Contrary to simple expectations that the two-dimensionalization enhances fluctuations, we observe that antiferromagnetic fluctuations are rapidly suppressed with decreasing Ce-BL thickness. 
Moreover, the suppression is more prominent near the interfaces between the BLs.  These results imply significant effects of local inversion-symmetry breaking at the interfaces.
\end{abstract}

\maketitle

The physics of materials with strong electron correlations is remarkably rich, and in these materials the entanglement of charge, spin and orbital degrees of freedom often leads to the emergence of exotic quantum phases.  
It has been shown that in the presence of strong spin-orbit coupling, the introduction of broken spatial inversion symmetry can produce further notable effects on these systems through a splitting of the Fermi surface with different spin structures even without a magnetic field, giving rise to a plethora  of novel phenomena such as anomalous magnetoelectric effects\cite{FujimotoJPSJ2007, BauerSpringer2012} and topological superconducting states.\cite{SatoPRL2009,SauPRL2010,LutchynPRL2010}
Moreover, it has been shown recently that the local inversion symmetry breaking at the interface of different compounds also dramatically affects the electronic properties, even when the global inversion symmetry is preserved in the whole crystals.\cite{HwangNatMat}   
Among others, the metallic state with the strongest electron correlation effects is achieved in the heavy-fermion $f$-electron systems, such as rare-earth and actinide compounds.
Recent advances in fabricating epitaxial superlattices consisting of heavy-fermion block layers (BLs) and conventional-metal BLs  \cite{ShishidoScience, MizukamiNatPhys} provide unique opportunity to study the effect of locally broken inversion symmetry at the interfaces between BLs.  

CeCoIn$_5$ [FIG.\;1(a)], one of the constituent of the CeCoIn$_5$/YbCoIn$_5$ superlattices studied here, possesses the highest superconducting transition temperature ($T_\mathrm{c} = 2.3$\,K) among Ce-based heavy-fermion superconductors,\cite{Ce115SC} in which an extremely narrow conduction band with effective mass (for about 100 times or more the bare electron mass)  is formed at low temperatures through the Kondo effect.  
CeCoIn$_5$ shares several common features with high-$T_\mathrm{c}$ cuprates, including unconventional superconducting gap with  $d_{x^2-y^2}$-symmetry \cite{AllanNatPhys,ZhouNatPhys, StockPRL08, IzawaPRL} and non-Fermi liquid properties.\cite{NakajimaJPSJ}
In addition, it is believed that the superconductivity is mediated by antiferromagnetic (AFM) fluctuations.  
In CeCoIn$_5$/YbCoIn$_5$ superlattices, several highly unusual superconducting properties have been observed, in particular when the thickness of the Ce-BLs is only a few unit cells.\cite{SweePRL, ShimozawaPRL}  
Although the importance of  the interfaces between Ce- and Yb-BLs has been emphasized, the lack of spectroscopic information prevents us from understanding their physical properties at the microscopic level.  
Thus the electronic and magnetic structures at the interfaces remain largely unexplored. 
\begin{figure}[tbhp]
\includegraphics[width=1\linewidth]{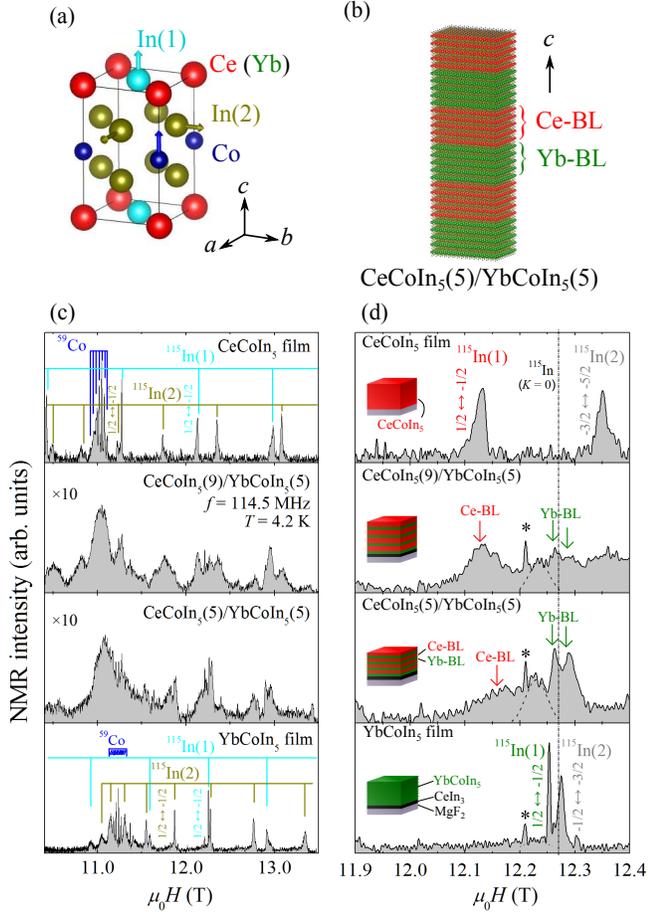}

\caption{
(color online). (a) Crystal structure of Ce(Yb)CoIn$_5$. The largest principle axes of the electric field gradient at the Co, In(1) and In(2) sites are shown by the arrows. (b) Schematic representation of the CeCoIn$_5$(5)/YbCoIn$_5$(5) superlattice.  (c)(d)  Field-sweep NMR spectra at 4.2\,K on thin-film CeCoIn$_5$ with the thickness of 500 nm, CeCoIn$_5$(9)/YbCoIn$_5$(5) and  CeCoIn$_5$(5)/YbCoIn$_5$(5) superlattices, and thin-film YbCoIn$_5$ with the thickness of 350 nm. Panels of (c) show the field range between 10.4 and 13.5\;T. 
Panels of (d) are the expanded views near the central transitions of the In(1) site. 
}
\end{figure}

Nuclear magnetic resonance (NMR) appears to be a particularly powerful probe, providing spatially-resolved microscopic information on the magnetic properties. 
We have performed NMR measurements on CeCoIn$_5$($n$)/YbCoIn$_5$(5) superlattices grown by molecular beam epitaxial technique,\cite{MizukamiNatPhys} where $n$ ($= 5$ or 9) layers of CeCoIn$_5$ and 5 layers of YbCoIn$_5$ were stacked alternately as shown in FIG.\;1(b).  Here the set of CeCoIn${_5} (n)$ and YbCoIn${_5} (5)$ repeats 40 times for $n = 5$ and 30 times for $n =9$.
Panels of FIG.\;1(c) depict the NMR spectra of the $n = 9$ and $n = 5$ superlattices for $\bm{H}\parallel c$, along with the spectra of CeCoIn$_5$ (Ce) [top] and YbCoIn$_5$ (Yb) [bottom] thin films.
The thickness of the Ce and Yb thin films is 500 nm and 350 nm, respectively.

From detailed analysis of the field dependence of the spectra we are able to obtain the site-selective NMR information, i.e. spectroscopic information resolved for Ce- and Yb-BLs separately.
In the thin films, NMR signals arising from two In sites, the In(1) located at the center of the Ce/Yb-In layer and the In(2) site  located on the lateral faces,  and the Co site can be clearly identified. 
The largest principal axis of the electric field gradient is parallel to the $c$ axis at the In(1)  and Co sites,  while it is perpendicular at the In(2) site. 
The parameters for the NMR line fitting are listed in the Table\;1, along with those of the bulk CeCoIn$_5$. 

\begin{table}
\caption{ NMR parameters of CeCoIn$_5$ and YbCoIn$_5$.
Knight shift along the $c$ axis, the electric quadrupole frequency and asymmetry parameter of the electric quadrupole interaction at Co, In(1), and In(2) sites in bulk CeCoIn$_5$, thin-film CeCoIn$_5$, and thin-film YbCoIn$_5$. The present parameters for bulk CeCoIn$_5$ are approximately the same values reported by Curro {\it et al.},\cite{Curro_shift,CurroNJP} and the data of thin-film samples are determined from the NMR spectra shown in FIG.~1.}
\begin{tabular}{c|c|ccc}
nuclear &sample &  $K_{\rm c}$(\%) & $\nu_Q$(MHz) & $\eta$ \\ \hline
$^{59}$Co    & CeCoIn$_5$(Bulk)    &3.39&0.230 &0  \\
             & CeCoIn$_5$(500\,nm) &3.58&0.302&0 \\
             & YbCoIn$_5$(350\,nm) &1.58&0.285&0 \\ \hline
$^{115}$In(1)& CeCoIn$_5$(Bulk)    &1.12&8.15&0  \\
             & CeCoIn$_5$(500\,nm) &1.14&8.04&0  \\
             & YbCoIn$_5$(350\,nm) &0.18&6.22&0  \\ \hline
$^{115}$In(2)& CeCoIn$_5$(Bulk)    &5.16&15.7&0.404\\
             & CeCoIn$_5$(500\,nm) &5.55&15.5&0.375 \\
             & YbCoIn$_5$(350\,nm) &0.40&14.0&0.490 \\ \hline
\end{tabular}
\end{table}

Here we focus on the spectra arising from the central transition ($+1/2 \leftrightarrow -1/2$) of the In(1) site, which range from 11.9 T to 12.4 T, as shown in panels of FIG.\;1(d).  
It should be noted that the spectra of the $n = 9$ and $n = 5$ superlattices consist of the spectra of Ce- and Yb-BLs.  
In fact, the positions of the spectra pointed by red and green arrows almost coincide with the peaks observed in the Ce and  Yb thin films, respectively.   
The maximum at around 12.22\;T observed in the $n = 5$ and 9 superlattices and the Yb thin film (asterisk and dashed peaks) arises from the CeIn$_3$ buffer layer.   
Owing to the zero asymmetric parameter at the In(1) site, central-transition peak ($+1/2 \leftrightarrow -1/2$)  of the In(1) site is not shifted by the electric quadrupole interaction, but shifted by the hyperfine interaction related to the local spin susceptibility.\cite{Metallic_shift}  
The Knight shift $K(T)$ at the fixed frequency $\omega_0$ is defined as 
\[K(T)=\left(\frac{H_0-H_{\rm res}}{H_{\rm res}}\right)_{\omega=\omega_0} =A_{\rm hf} \chi(T) \]
where $H_{\rm res}$ and $H_0$ are resonant magnetic fields of a sample and a bare nucleus that has the relation of $\omega_0 = \gamma_n H_0$ with the nuclear gyromagnetic ratio $\gamma_n$.  
$A_{\rm hf}$ and $\chi(T)$ are the hyperfine coupling constant and the local spin susceptibility, respectively.  
The shift from $\mu_0H_0=12.27$\;T ($K = 0$) is proportional to the local static susceptibility at the In(1) site. 

The nuclear spin-lattice relaxation rate $T_1^{-1}$ provides microscopic information on the dynamical  magnetic properties; $(T_1T)^{-1}$ is proportional to the momentum {\boldmath $q$}-summed imaginary part of the dynamical susceptibility, i.e. $(T_1T)^{-1} \propto \sum_{q }A_{\rm hf}^2 \chi ^{\prime \prime}(q, \omega)/\omega$.   
Figure\;2 depicts  $(T_1T)^{-1}$ at the In(1) site of two thin films and superlattices with $n$ = 5 and 9  at $\mu_0H \sim 12$\;T.    
Although the In(1) site is located at the symmetric position in the unit cell, In(1) sees the AFM fluctuations from rare earth ions since the bond axes of the ordered moments are not coincident with the unit cell;\cite{CurroNJP}  $(T_1T)^{-1}$ is dominated by $\chi(\bm{Q})$ with  $\bm{Q}=(\pi/a, \pi/a, \pi/c)$, i.e. AFM $\bm{Q}$ vectors.\cite{StockPRL08}
In the Yb thin film, $(T_1T)^{-1}$ is temperature independent (Korringa relation), which is typical behavior of the uncorrelated nonmagnetic metal. 
In the Ce thin film, on the other hand, $(T_1T)^{-1}$ is strongly enhanced at low temperatures, indicating the presence of strong AFM fluctuations. 
In FIG.\;2, dashed line represents $(T_1T)^{-1}$ of the CeCoIn$_5$ single crystal measured at the same field, indicating that the AFM fluctuations in the Ce thin film are the same as those in bulk single crystals.

\begin{figure}[tbhp]
\includegraphics[width=0.8\linewidth]{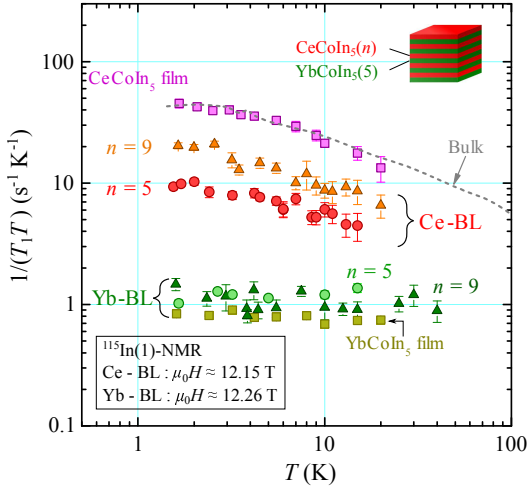}
\caption{
(color online). The nuclear spin-lattice relaxation rate divided by temperature $(T_1T)^{-1}$ at the In(1) site in thin-film CeCoIn$_5$ (magenta squares), thin-film YbCoIn$_5$ (khaki squares) and CeCoIn$_5(n)$/YbCoIn$_5$(5) superlattices with $n = 9$ (triangles) and $n = 5$ (circles). In the superlattices, $(T_1T)^{-1}$s in the Ce-BLs (orange and red) and Yb-BLs (dark and light green) are shown separately. 
$(T_1T)^{-1}$ in bulk CeCoIn$_5$ measured at $\mu_0 H \sim 12.1$ T is also shown with dashed line. 
} 
\end{figure}

Figures 3(a) and (b) show the expanded NMR spectra at 3.2\,K near the In(1) central transition of the $n = 9$ and $n = 5$ superlattices. 
The NMR spectra of these superlattices contain signals from three different layers, i.e. Ce- and Yb-BLs  and CeIn$_3$ buffer layers, each of which has different Knight shift.  
It should be stressed that the spectra from each layer are well separated in this field range. 
In FIGs.\;3(a) and (b), the red, orange, and yellow shaded regions represent the spectra of Ce-BLs, the green region represent the spectra of Yb-BLs and the peak shown by asterisk represent the spectra of CeIn$_3$ buffer layers.  
These assignments are made by the straightforward comparison with the spectra of CeCoIn$_5$ and YbCoIn$_5$ thin films. 
A salient feature is that the shape of the spectra of Ce-BLs for $n = 9$ is different from that of $n = 5$. 
The spectrum in the lower-field region of $n = 9$ has a much larger weight compared with that of $n = 5$.
This naturally implies that the spectra in higher field regimes shaded by yellow arise from the outer CeCoIn$_5$ layers close to the interfaces (FIG.\;3(e)), because the fraction of the interface layers increases rapidly with the reduction of $n$.
In FIGs. \;3(a) and (b), the red to yellow gradation is a schematic representation of the NMR signal arising from a inner and outer (interface) layers in Ce-BLs.
The colors of the spectra correspond to the colors in the Ce-BL in FIG. 3(e).
Thus the field dependence of $(T_1T)^{-1}$  enables us to resolve the layer dependence of the magnetic fluctuations even within a same Ce-BLs. 
The field dependence of $(T_1T)^{-1}$ shown in FIG.\;3(c) indicates that even within the Ce-BLs AFM fluctuations have strong spatial dependence; AFM fluctuations near the interface are weaker than those at the inner Ce-layers. 
This indicates that the suppression of the AFM fluctuations is larger near the interface than in the inner layers.
The field dependence of $1/T_1$s in FIG. 3(c) is clearly recognized from the recovery of the nuclear magnetization $m(t)$ at a time $t$ after a saturation pulse(FIG.3 (d)). The recovery $\left( m(\infty ) -m(t)\right) /m(\infty)$ at 12.180 T is more slowly damped than the recovery at 12.115 T.

\begin{figure}[tb]
\includegraphics[width=1\linewidth]{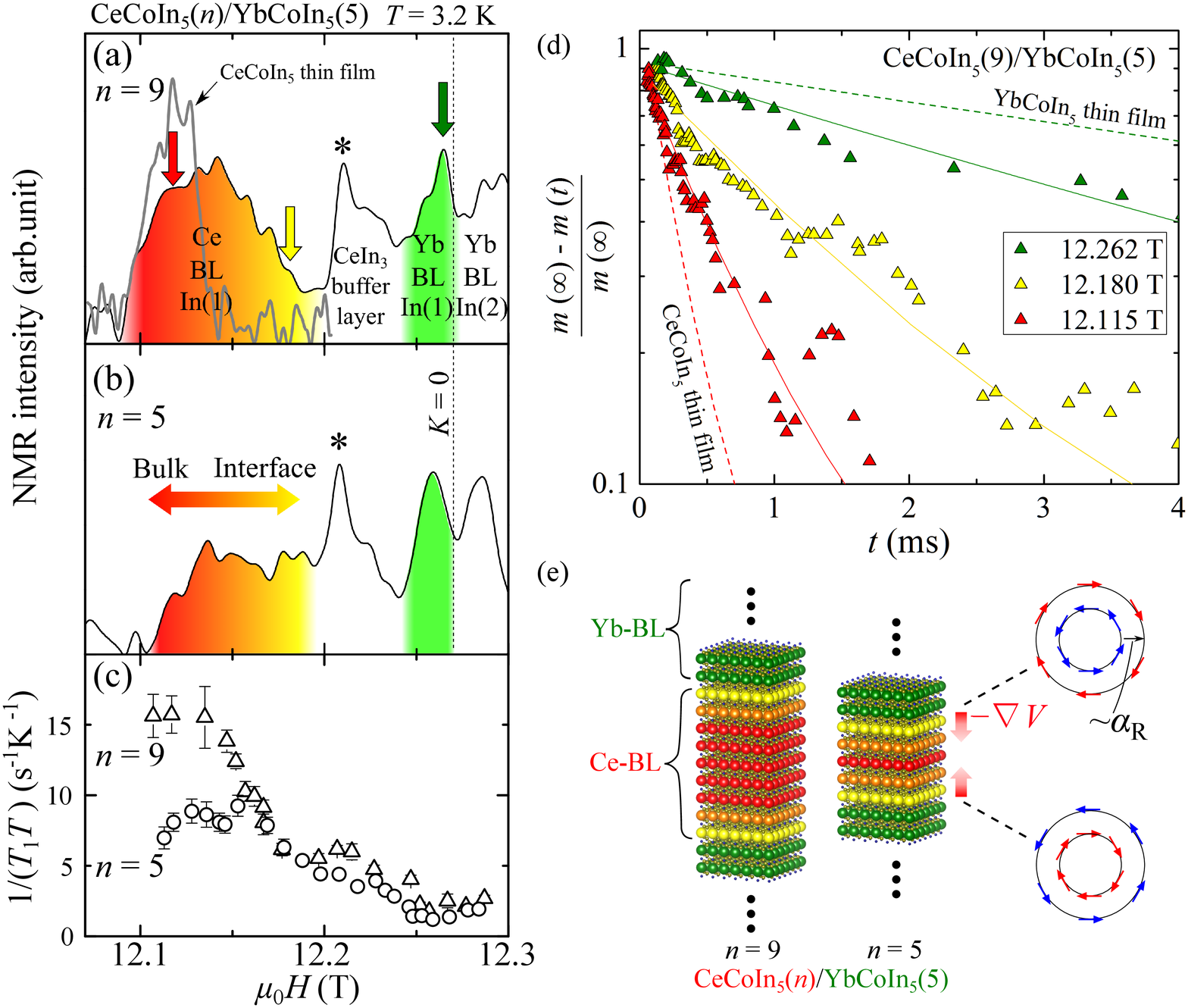}
\caption{
(color online). (a), (b) $^{115}$In(1)-NMR spectra of the $n = 9$ (a) and $5$  (b) superlattices at 3.2\,K and with fixed NMR frequency ($f =114.5$\,MHz). 
In (a), the red(12.115 T), yerrow(12.180 T) and green(12.262 T) arrow shows the magnetic field at which we observed relaxation curves. 
(c) Field dependence of $(T_1T)^{-1}$ in the $n = 9$ and $n = 5$ superlattices. (d) Relaxation curves vs $t$ plot at 12.262 T, 12,115 T and 12.180 T with $f =114.5$\,MHz in CeCoIn$_5$(9)/YbCoIn$_5$(5). The red(green) dashed line represents the fitted relaxation curve of CeCoIn$_5$(YbCoIn$_5$) thin film.
(e)  The schematic Ce and Yb arrangements of these superlattices. From the comparison of two $^{115}$In(1)-NMR spectra, the field-dependent NMR signals are approximately assigned to arise from the colored layers in (e).  
} 
\end{figure}

The temperature dependence of $(T_1T)^{-1}$ of the Ce- and Yb-BLs in each superlattice is plotted separately in FIG.\;2. 
Here the relaxation rate of Ce-BLs is determined at the maximum of the broad peak, which represents an average of the AFM fluctuations in the whole Ce-BLs. 
It is clear that while relaxation rate in the Yb-BLs is essentially unchanged, but the magnitude of $(T_1T)^{-1}$ in the Ce-BLs is suppressed as the Ce-BLs thickness is reduced. 
We analyze  $(T_1T)^{-1}$ in terms of the Curie-Weiss (CW) formula [$(T_1T)^{-1} \propto C/(T+\theta)$], assuming that  dynamical susceptibility is dominated by the AFM fluctuations. 
Here $\theta$ is the Weiss temperature and $C$ is related to the Curie constant.  
The experimental data can be fitted by the CW formula as shown in FIG.\;4.  
The reduction of $n$ leads to a serious reduction of $C$, indicating that the AFM fluctuations are suppressed as the Ce-BLs become thinner.   
We emphasize that the proximity of $f$-electrons with magnetic moment to nonmagnetic Yb-layers is unlikely to be the origin of this reduction, because of the following reasons.  
First,  magnetic fluctuations in the Yb-BLs are essentially the same as those in the YbCoIn$_5$ thin film (FIG.\;2).  
This indicates that the magnetic degrees of freedom of $f$-electrons are quenched inside the Yb-BLs. Second, it has been shown that the inter-diffusion between Ce- and Yb-BLs is negligibly small,\cite{MizukamiNatPhys} and that large Fermi velocity mismatch across the interface leads to huge suppression of the transmission probability of $f$-electrons.\cite{ShePRL2012}  
Third, as reported experimentally,\cite{MizukamiNatPhys, SweePRL, ShimozawaPRL} while $T_\mathrm{c}$ in these superlattices is suppressed from $T_\mathrm{c}$ in the bulk, upper critical field $H_{\mathrm{c}2}$ are enhanced from the bulk value for both $\bm{H}\parallel ab$ and $c$.  
This enhancement of $H_{\mathrm{c}2}$ is qualitatively inconsistent with the proximity of $f$-electrons. 

\begin{figure}[tb]
\includegraphics[width=0.8\linewidth]{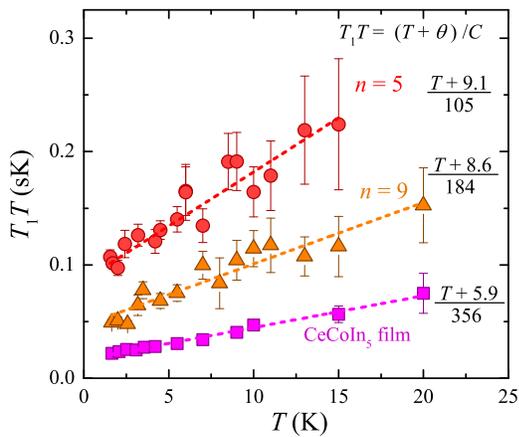}
\caption{
(color online). Plot of $T_1T$ at the In(1) site of Ce-BLs against $T$ in the $n = 9$ and $n = 5$ superlattices compared with the result of pure CeCoIn$_5$ films. The data can be fitted by the Curie-Weiss law [$T_1T = (T+\theta)/C$].
} 
\end{figure}

The observed suppression of the magnetic fluctuations in the Ce-BLs is opposite to the enhancement of fluctuations expected from two-dimensionalization, which leads to the enhancement of the density of states at the Fermi level.  
The spatial dependence of the relaxation rate within the Ce-BLs [FIG.\;3(c)] strongly suggests that the breaking of the local inversion symmetry at the interfaces between Ce- and Yb-BLs plays a decisive role for determining the magnetic properties of the Ce-BLs, in particular when their thickness is only a few unit-cell thick.   
In the absence of inversion symmetry, an asymmetric potential gradient $\nabla V$ yields a spin-orbit interaction.  
When $\nabla V$ is perpendicular to the two-dimensional plane, $\nabla V \parallel c$, Rashba spin-orbit  interaction $\alpha_{\rm R}\bm{g}(\bm{k})\cdot\bm{\sigma} \propto (\bm{k}\times \nabla V) \cdot \bm{\sigma}$ splits the Fermi surface into two sheets with different spin structures, where $\bm{g}(\bm{k})=(-k_y, k_x, 0)/k_{\rm F}$, $k_{\rm F}$ is the Fermi wave number, and $\bm{\sigma}$  is the Pauli matrix.
The energy splitting is given by $\alpha_{\rm R}$, and this interaction locks spin configurations within the $ab$ plane with clockwise rotation on one sheet and anticlockwise on the other, as shown in FIG.\;3(e).\cite{FujimotoJPSJ07}  
Here, the inversion symmetry is {\it locally} broken at the top and the bottom layers of the Ce-BLs at the immediate proximity to the Yb-BLs. 
In the presence of the local inversion symmetry breaking together with the fact that the Ce atom has a large atomic number, the Rashba-type spin-orbit coupling is expected to be strong in the present superlattices.   
The importance of the local inversion symmetry breaking at the interface has been emphasized experimentally through the peculiar angular variation of upper critical field, which can be interpreted as a strong suppression of the Pauli pair-breaking effect.\cite{SweePRL} 
The Fermi surface splitting due to the Rashba coupling should modify seriously the nesting condition and hence is expected to reduce the commensurate AFM fluctuations with $\bm{Q}=(\pi/a,\pi/a,\pi/c)$, which is dominant in bulk CeCoIn$_5$.   
In addition, it has been pointed out that the broken inversion symmetry at the interface reduces the AFM fluctuations by lifting the degeneracy of the fluctuation modes through the helical anisotropy of the spin configuration shown in FIG.\;3(e).\cite{YanaseJPSJ08, TakimotoJPSJ08}  
In fact, similar situation has been reported in ferromagnetic metal MnSi, where helical  spin structure is stabilized by the spin-orbit coupling.\cite{NakanishiSSC80}
With the reduction of $n$, the fraction of the noncentrosymmetric interface layers increases rapidly, leading to the suppression of the AFM fluctuations.  
The present results suggest that the local inversion symmetry breaking plays a key role for the magnetic properties at the interface of strongly correlated electron systems, which is expected to host a fertile ground for observing exotic properties.\cite{YoshidaJPSJ13}

The authors thank S. K. Goh, R. Peters, Y. Tada, Y. Yanase and H. Ikeda for valuable discussions. 
This work was partially supported by Kyoto Univ. LTM center, Grant-in-Aid from the Ministry of Education, Culture, Sports, Science, Technology(MEXT) of Japan, Grants-in-Aid for Scientific Research (KAKENHI) from Japan Society for the Promotion of Science (JSPS), ``Topological Quantum Phenomena'' (No.\,25103713) Grant-in-Aid for Scientific Research on Innovative Areas from the Ministry of Education, Culture, Sports, Science, and Technology (MEXT) of Japan. Figures of crystal structures were shown using the VESTA package.\cite{vesta}


%

\end{document}